\documentstyle[11pt]{article}

\def\fnote#1#2{\begingroup\def\thefootnote{#1}\footnote{#2}\addtocounter
{footnote}{-1}\endgroup}

\begin{document}

\hfill{UTTG-01-08}

\vspace{36pt}

\begin{center}
{\large {\bf {Effective Field Theory for Inflation}}}

\vspace{36pt}
Steven Weinberg\fnote{*}{Electronic address:
weinberg@physics.utexas.edu}\\
{\em Theory Group, Department of Physics, University of
Texas\\
Austin, TX, 78712}

\vspace{30pt}

\noindent
{\bf Abstract}
\end{center}
\noindent
The methods of effective field theory are used to study generic theories of inflation with a single inflaton field.  For scalar modes, the leading corrections to the ${\cal R}$ correlation function are found to be purely of the $k$-inflation type.  For tensor modes the leading corrections to the correlation function arise from terms in the action that are quadratic in the curvature, including a parity-violating term that makes the propagation of these modes depend on their helicity.  These methods are also briefly applied to non-generic theories of inflation with an extra shift symmetry, as in so-called ghost inflation.

 \vfill

\pagebreak

\begin{center}
{\bf I. Generic Theories of Inflation}
\end{center}

Observations of the cosmic microwave background and large scale structure are consistent with a simple theory of inflation[1] with a single canonically normalized inflaton field $\varphi_c(x)$, described by a Lagrangian
\begin{equation}
{\cal L}_0=\sqrt{g}\Bigg[-\frac{M_P^2}{2}R-\frac{1}{2}g^{\mu\nu}\partial_\mu\varphi_c \partial_\nu\varphi_c-V(\varphi_c)\Bigg]\;,
\end{equation}
where $g\equiv -{\rm Det} g_{\mu\nu}$, $M_P\equiv 1/\sqrt{8\pi G}$ is the reduced Planck mass, and $V(\varphi_c)$ is a potential down which the scalar field  rolls more-or-less slowly.    With this theory, the strength of observed fluctuations in the microwave background matter density indicates that the cosmic expansion rate $H\equiv \dot{a}/a$ and the physical wave number $k/a$ at horizon exit, when these are equal,  have the value[2] $H=k/a\approx \sqrt{\epsilon}\times 2\times 10^{14}$ GeV, where $\epsilon$ is the value of $-\dot{H}/H^2$ at this time, and $a$ is the Robertson--Walker scale factor.  Hence $H$ and $k/a$ at horizon exit are likely to be much less than $M_P\simeq 2.4\times 10^{18}$ GeV, and even considerably less than a plausible grand unification scale $\approx 10^{16}$ GeV.  This provides a justification after the fact for using a Lagrangian (1) with a minimum number of spacetime derivatives.  (As is well known, (1) is the most general Lagrangian density for gravitation and a single scalar field with no more than two spacetime derivatives.  An arbitrary  function of $\varphi$ multiplying the first term could be eliminated by a redefinition of the metric, and an arbitrary  function of $\varphi$ multiplying the second term could be eliminated by a redefinition of $\varphi$.)

But $H$ and $k/a$ at horizon exit are not entirely negligible compared with whatever fundamental scale  characterizes the theory underlying inflation, and at earlier times $k/a$ is exponentially larger than at horizon exit, so it is worth considering the next corrections to (1).  We assume that (1) is just the first term in a generic effective field theory, in which terms with higher derivatives are suppressed by negative powers of some large mass $M$, characterizing whatever fundamental theory underlies this effective field theory.  Rather than committing ourselves to any particular underlying theory, we will simply assume that all constants in the higher derivative terms of the effective Lagrangian take values that are  powers of $M$ indicated by dimensional analysis, with coefficients roughly of order unity.  Because $H$ and $k/a$ are so large during inflation, observations of fluctuations produced during inflation provide a  unique opportunity for detecting effects of  higher derivative terms in the gravitational action.  

To get some idea of the value of $M$, we note that the unperturbed canonically normalized scalar field $\bar{\varphi}_c$ described by the Lagrangian (1) has a time derivative $\dot{\bar{\varphi}}_c=\sqrt{2\epsilon}M_PH$, so the change in $\bar{\varphi}_c$ during a Hubble time $1/H$ at around the time of horizon exit is of order $\dot{\bar{\varphi}}_c/H=\sqrt{2\epsilon}M_P$.  If we are to use effective field theory to study fluctuations at about the time of horizon exit in generic theories in which the dependence of the action on $\varphi_c$ is unconstrained by symmetry principles or by other consequences of an  underlying theory, and if (1) is at least a fair first approximation to the full theory, then the  mass $M$ that is characteristic of the effective field theory of inflation can not be much smaller than $\sqrt{2\epsilon}M_P$, for if it were then there would be no limit on the size of  higher-derivative terms containing many powers of $\varphi_c/M$.    It follows that the expansion parameter $H/M$ in this class of theories is no greater than $H/\sqrt{2\epsilon}M_P\simeq 6\times 10^{-5}$, whatever the value of $\epsilon$.

We will tentatively assume here that $M$ is of order $\sqrt{2\epsilon}M_P$, in which 
case the coefficients of the higher-derivative terms in the effective Lagrangian have to be taken as arbitrary functions of $\varphi_c/M$.  This is likely to be the case if $\epsilon$ is not too small, say of order 0.02, since then there is not much
 difference between $\sqrt{2\epsilon}M_P$ and $M_P$, and $M$ is unlikely to be much larger than $M_P$.  (The considerations presented below would still be valid if $M$ were instead much larger than $\sqrt{2\epsilon}M_P$, as for instance if $M\approx M_P$ and $\epsilon$ is very small, but then we would have to count powers of $\varphi_c/M$ as well as numbers of derivatives in judging how much the various 
higher-derivative terms are suppressed, and some of the coefficient functions in the effective Lagrangian derived below would be negligible, and all others very simple.)   
From now on we will work with a dimensionless scalar field $\varphi\equiv \varphi_c/M$, and write (1) as
\begin{equation}
{\cal L}_0=\sqrt{g}\Bigg[-\frac{M_P^2}{2}R-\frac{M^2}{2}g^{\mu\nu}\partial_\mu\varphi \partial_\nu\varphi-M_P^2 U(\varphi)\Bigg]\;,
\end{equation}
where  $U(\varphi)\equiv V(M\varphi)/M_P^2$.  Note that the unperturbed value of $U$ is $(3-\epsilon)H^2$, so we can think of $U$ as well as  $\partial_\mu\varphi\partial^\mu\varphi$ as both being of order $H^2$ at horizon exit.

  The leading correction to (2) will  consist of a sum of all generally covariant terms with four spacetime derivatives and coefficients of order unity[3].  By a judicious weeding out of  of total derivatives, the  most general such correction term   can be put in the form[4] 
\begin{eqnarray}
\Delta{\cal L}&=&\sqrt{g}\Bigg[f_1(\varphi)\Big(g^{\mu\nu}\varphi_{,\mu}\varphi_{,\nu}\Big)^2
+f_2(\varphi)g^{\rho\sigma}\varphi_{,\rho}\varphi_{,\sigma}\Box\varphi\nonumber\\&&
+
f_3(\varphi)\Big(\Box \varphi\Big)^2
+f_4(\varphi)R^{\mu\nu}\varphi_{,\mu}\varphi_{,\nu}
+f_5(\varphi)R\, g^{\mu\nu}\varphi_{,\mu}\varphi_{,\nu}\nonumber\\&&
+f_6(\varphi)R\, \Box \varphi
+f_7(\varphi)R^2\nonumber\\&&+f_8(\varphi)R^{\mu\nu}R_{\mu\nu}
+f_9(\varphi)C^{\mu\nu\rho\sigma}C_{\mu\nu\rho\sigma}\Bigg]\nonumber\\&&+f_{10}(\varphi)\epsilon^{\mu\nu\rho\sigma}C_{\mu\nu}{}^{\kappa\lambda}C_{\rho\sigma\kappa\lambda}\;,
\end{eqnarray}
where  as usual commas denote ordinary derivatives and semicolons denote covariant derivatives; $\Box\varphi\equiv g^{\mu\nu}\varphi_{,\mu;\nu}$ is the invariant d'Alembertian of $\varphi$;   $\epsilon^{\mu\nu\rho\sigma}$ is the totally antisymmetric tensor density with $\epsilon^{1230}\equiv +1$;  and the $f_n(\varphi)$ are dimensionless functions, treated here as of order unity.  In the last two terms, instead of the Riemann--Christoffel tensor $R_{\mu\nu\rho\sigma}$, we have used the Weyl tensor
\begin{equation}
C_{\mu\nu\rho\sigma}\equiv R_{\mu\nu\rho\sigma}-\frac{1}{2}\Big(g_{\mu\rho}R_{\nu\sigma}-g_{\mu\sigma}R_{\nu\rho}
-g_{\nu\rho}R_{\mu\sigma}+g_{\nu\sigma}R_{\mu\rho}\Big)+\frac{R}{6}\Big(g_{\mu\rho}g_{\nu\sigma}-g_{\nu\rho}g_{\mu\sigma}\Big)\;.
\end{equation}
Writing the last two terms in Eq.~(3) as  bilinears in $C_{\mu\nu\rho\sigma}$ rather than $R_{\mu\nu\rho\sigma}$ has no effect in the last term, and in the penultimate term of course just amounts to a different definition of $f_7$ and $f_8$.  (Similarly,  instead of writing the penultimate term as a bilinear in $C_{\mu\nu\rho\sigma}$ or $R_{\mu\nu\rho\sigma}$, we could have written it as the linear combination of curvature bilinears that appears in the Gauss--Bonnet identity; even though this linear combination is a total derivative, it would affect the field equations because its coefficient $f_9(\varphi)$ is not constant.)  Our reason for choosing to use the Weyl tensor in the last two terms will become apparent soon.

The correction term (3) involves second time derivatives, as well as fields and their first time derivatives.   If we took such a theory literally, we would find more than just the usual two adiabatic modes for single-field inflation, and the commutation relations (as given by the Ostrogradski formalism[5]) would be bizarre, with $\varphi$ commuting with $\dot{\varphi}$.  (For instance, Kallosh, Kang, Linde, and Mukhanov[6] encounter such additional modes when the Ostrogradski formalism is applied to a scalar field Lagrangian involving second time derivatives.)  Similarly, there are metric components (such as $g^{00}$ and $g^{0i}$ in the ADM formalism[7]) whose time derivatives do not appear in ${\cal L}_0$, but that do appear in $\Delta{\cal L}$.  If we were to take ${\cal L}_0+\Delta{\cal L}$  as the full Lagrangian, then the correction term $\Delta{\cal L}$ would cause these auxiliary fields to become dynamical, with a further expansion of the modes of the system.  

Instead, we should remember that from the point of view of effective field theory, Eqs.~(2) and (3)  represent just the lowest two terms in an expansion in inverse powers of $M$, so we must rule out any modes that cannot be expanded in this way[8].  This means in particular that we {\em must}  eliminate all second time derivatives and time derivatives of auxiliary fields in the first correction terms in the effective action by using the field equations derived from the leading terms in the action.\footnote{This is equivalent to what is generally done in deriving Feynman rules in effective flat-space quantum field theories.  Consider for instance the very simple effective Lagrangian 
$${\cal L}=-\frac{1}{2}[\partial_\mu\varphi\partial^\mu\varphi+m^2\varphi^2+M^{-2}(\Box\varphi)^2]+J\varphi\,$$
where $M\gg m$ is some very large mass, and $J$ is a c-number external current.  We can easily find the connected part  $\Gamma$ of the vacuum persistance amplitude:
$$
\Gamma=i\int d^4k \frac{|J(k)|^2}{k^2+m^2+k^4/M^2}\;.
$$
If we took this result seriously, then we would conclude that in addition to the usual particle with mass $m+O(m^3/M^2)$, the theory contains an unphysical one particle state with  mass $M+O(m^2/M)$.  But if we regard ${\cal L}$ as just the first two terms in a power series in $1/M^2$, then we must treat the term $M^{-2}(\Box\varphi)^2$ as a first-order perturbation, so that the vacuum persistence amplitude is 
$$
\Gamma=i\int d^4k |J(k)|^2\left[\frac{1}{k^2+m^2}-\frac{k^4}{M^2(k^2+m^2)^2}+\dots\right]\;,
$$
and the only pole is at  $k^2=-m^2$.
This is just the same result for $\Gamma$ that we would find if we were to eliminate the second time derivatives in the $O(M^{-2})$ term in $\cal L$ by using the field equation derived from the leading term in the Lagrangian
$$\Box\varphi=m^2\varphi-J\;.$$
In this case the effective Lagrangian becomes 
$${\cal L}=-\frac{1}{2}[\partial_\mu\varphi\partial^\mu\varphi+m^2\varphi^2+m^4M^{-2}\varphi^2]+(1+m^2/M^2)J\varphi-J^2/2M^2\;.$$
Taking into account all $J$-dependent terms, it is straightforward to see that with this Lagrangian we get the same vacuum persistence amplitude as found above for the  the original Lagrangian, when  $M^{-2}(\Box\varphi)^2$ is treated as a first-order perturbation.}
  In the present case,  we must eliminate second time derivatives and time derivatives of auxiliary fields in (3) by using the zero-th order field equations derived from (2):
\begin{equation}
 M^2\Box\varphi=M_P^2 U'(\varphi)\;,~~~~~~R_{\mu\nu}=-(M^2/M_P^2)\varphi_{,\mu}\varphi_{,\nu}-U(\varphi) g_{\mu\nu}\;.
\end{equation}

Using these field equations in Eq.~(3) allows us, with some redefinitions, to eliminate all of the terms in (3) except the first one and the last two.  Specifically, the second term in Eq.~(3)   just provides a field-dependent correction to the kinematic term in (2), which can be eliminated by a redefinition of the inflaton field; the third term just provides a correction $f_3{U'}^2M_P^4/M^4$ to the potential in (2), which can be absorbed into a redefinition of $U(\varphi)$; the fourth and fifth  terms  supply corrections to both $f_1(\varphi)$ and the kinematic term in (2); the sixth term provides corrections to the kinematic term and the potential in (2); and the seventh and eighth   terms provide corrections to the kinematic term and potential in (2) and to $f_1(\varphi)$.  That is, with suitable redefinitions of $\varphi$, $U(\varphi)$, and $f_1(\varphi)$, and with various total derivatives dropped, the Lagrangian is the sum of (2) and a correction term of the form 
\begin{equation}
\Delta{\cal L}=\sqrt{g}f_1(\varphi)\Big(g^{\mu\nu}\varphi_{,\mu}\varphi_{,\nu}\Big)^2
+\sqrt{g}f_9(\varphi)C^{\mu\nu\rho\sigma}C_{\mu\nu\rho\sigma}+f_{10}(\varphi)\epsilon^{\mu\nu\rho\sigma}
C_{\mu\nu}{}^{\kappa\lambda}C_{\rho\sigma\kappa\lambda}\;,
\end{equation}
The first term is of the type encountered in theories of ``$k$-inflation''[9].  This term must be included in the Lagrangian, as a counterterm to ultraviolet divergences encountered when the leading terms in (2) are used in one-loop order.  The second term (or an equivalent Gauss--Bonnet term)  has been considered in connection with inflation and the evolution of dark energy[10].  

For a general function $f_{10}(\varphi)$ the final term in Eq.~(6) violates parity conservation[11].  That is, although the action is invariant under coordinate transformations $x^\mu\rightarrow x'^\mu$ that are ``small,'' in the sense that ${\rm Det}(\partial x'/\partial x)>0$, it is not invariant under {\em inversions}, that is, under coordinate transformations with ${\rm Det}(\partial x'/\partial x)<0$.  It is only invariance under ``small'' coordinate transformations that is needed to ensure the conservation of the energy-momentum tensor, and no sequence of ``small'' coordinate transformations can ever add up to an inversion, so there is no {\em a priori}  reason to impose invariance under inversions, including  space inversion.  The fact that parity has always been observed to be conserved in gravitational interactions is sufficiently explained by the fact that terms in the effective action for gravity and scalars  with no more than two spacetime derivatives that are invariant under ``small'' coordinate transformations cannot be complicated enough to violate invariance under inversions.  

From now on we shall work  in  perturbation theory, writing
\begin{equation}
g_{\mu\nu}({\bf x},t)=\bar{g}_{\mu\nu}(t)+h_{\mu\nu}({\bf x},t) \;,~~
\varphi({\bf x},t)=\bar{\varphi}(t)+\delta\varphi({\bf x},t)\;,
\end{equation}
where  $\bar{g}_{\mu\nu}(t)$ is the flat-space Robertson--Walker metric with $\bar{g}_{00}=-1$, $\bar{g}_{0i}=0$, and $\bar{g}_{ij}=a^2(t)\delta_{ij}$; $\bar{\varphi}(t)$ is the unperturbed scalar field; and $h_{\mu\nu}$ and $\delta\varphi$ are first-order perturbations.  In this paper  we will mostly be concerned  with the terms in the Lagrangian  that are quadratic in perturbations, which are needed for the calculation of Gaussian correlations.  Terms of higher order in perturbations 
that are needed for the calculation of non-Gaussian effects  will be considered only briefly.  

Because the spatially flat 
Robertson--Walker metric is also conformally flat it has a vanishing Weyl tensor, and so  the Weyl tensor starts with a term of first order in perturbations.  This saves us from having to calculate the Weyl tensor to second order in perturbations; we have simply  
\begin{eqnarray}
&&[\Delta{\cal L}]^{(2)}=\left[\sqrt{g}f_1(\varphi)\Big(g^{\mu\nu}\varphi_{,\mu}\varphi_{,\nu}\Big)^2\right]^{(2)}
\nonumber\\&&+a^3f_9(\bar{\varphi})\bar{g}^{\mu\kappa}\bar{g}^{\nu\lambda}\bar{g}^{\rho\eta}\bar{g}^{\sigma\zeta}
C^{(1)}_{\kappa\lambda\eta\zeta}C^{(1)}_{\mu\nu\rho\sigma}+f_{10}(\bar{\varphi})\epsilon^{\mu\nu\rho\sigma}
\bar{g}^{\kappa\eta}\bar{g}^{\lambda\zeta}C^{(1)}_{\mu\nu\eta\zeta}C^{(1)}_{\rho\sigma\kappa\lambda}\;,~~~~~~
\end{eqnarray}
where the superscripts $(1)$ and $(2)$ denote terms of first and second order in perturbations, respectively.  Furthermore, the Weyl tensor is traceless, which to first order in perturbations gives
\begin{equation}
C^{(1)}_{i0k0}=a^{-2}C^{(1)}_{ijkj}\;,~~~C^{(1)}_{ijj0}=C^{(1)}_{i0i0}=C^{(1)}_{ijij}=0\;.
\end{equation}
Since scalar and tensor fluctuations do not interfere in Gaussian correlations, they will  be considered separately.

\begin{center}
{\bf  II. Scalar Fluctuations}
\end{center}

Here we are interested in terms in the Lagrangian that, after eliminating auxiliary fields, are quadratic in ${\cal R}$,  the familiar gauge-invariant quantity that is conserved outside the horizon[1]
\begin{equation}
{\cal R}\equiv \frac{A}{2}-\frac{H\delta\varphi}{\dot{\bar{\varphi}}}\;,
\end{equation}
with $A$  defined by writing the spatial part of the metric perturbation for scalar perturbations in a general gauge as 
\begin{equation}
h_{ij}({\bf x},t)=a^2(t)\left[\delta_{ij}A({\bf x},t)+\frac{\partial^2 B({\bf x},t)}{\partial x^i\partial x^j}\right]\;.
\end{equation} 
Let us consider in turn the contribution  of  the three terms in Eq.~(8) to the quadratic part of the Lagrangian for ${\cal R}$.

First, terms in the effective Lagrangian like the first term in Eq.~(8) that depend only on $\varphi$ and $\partial_\mu\varphi\partial^\mu\varphi$ are known to enter into the part of the Lagrangian quadratic in scalar fluctuations only through their effect on the sound speed $c_s(t)$[12]. That is, after eliminating auxiliary fields, 
\begin{equation}
\left[-\frac{\sqrt{g}M_P^2}{2}R+\sqrt{g}P\Big(-\partial_\mu\varphi\partial^\mu\varphi/2,\varphi\Big)\right]^{(2)}
=-\frac{M_P^2\dot{H}}{H^2}a^3\left[\frac{1}{c_s^2}\dot{{\cal R}}^2-\frac{1}{a^2}(\vec{\nabla}{\cal R})^2\right]\;,
\end{equation}
where
\begin{equation}
c_s^{-2}=1+2\left[X \frac{\partial^2P(X,\bar{\varphi})}{\partial X^2}\Bigg/\frac{\partial P(X,\bar{\varphi})}{\partial X}\right]_{X=\dot{\bar{\varphi}}^2/2}\;.
\end{equation}
In particular, the first term in Eq.~(8) shifts the squared speed of sound by
\begin{equation}
\Delta c_s^2=\frac{16\dot{H}M_P^2f_1(\bar{\varphi})}{M^4}\;,
\end{equation} 
corresponding to a second-order perturbation 
\begin{equation}
\left[\sqrt{g}f_1(\varphi)\Big(g^{\mu\nu}\varphi_{,\mu}\varphi_{,\nu}\Big)^2\right]^{(2)}=\frac{16 M_P^4 a^3\dot{H}^2f_1(\bar{\varphi})}{M^4H^2}\dot{{\cal R}}^2 \;.
\end{equation}

The other terms in Eq.~(8) are greatly simplified by noting that, for scalar modes, $C_{ijk0}$ must take the form of   $\delta_{ik}\partial_j-\delta_{jk}\partial_i$ acting on some scalar, so the above trace condition  $C^{(1)}_{ijj0}=0$ implies that $C^{(1)}_{ijk0}=0$.   Hence  all we need to evaluate the second term in (8) are the purely spatial components of the Weyl tensor.  Using the field equations (5) to eliminate auxiliary fields and second time derivatives, we find after a straightforward though tedious calculation that
\begin{eqnarray}
&& a^3f_9(\bar{\varphi})\bar{g}^{\mu\kappa}\bar{g}^{\nu\lambda}\bar{g}^{\rho\eta}\bar{g}^{\sigma\zeta}
C^{(1)}_{\kappa\lambda\eta\zeta}C^{(1)}_{\mu\nu\rho\sigma}=
a^{-5}f_9(\bar{\varphi})
\left[C^{(1)}_{ijkl}C^{(1)}_{ijkl}+4C^{(1)}_{ijkj}C^{(1)}_{ilkl}\right]\nonumber\\&&~~~~=
\frac{16\dot{H}^2}{3H^2}a^3f_9
(\bar{\varphi})\dot{{\cal R}}^2\;.
\end{eqnarray}
Comparing this with Eq.~(15), we see that the effect of the second term in Eq.~(8) is the same as a change in the coefficient $f_1$ of the first term by an amount
\begin{equation}
\Delta f_1(\varphi)=\frac{M^4}{3M_P^4}f_9(\varphi)\;.
\end{equation}

Finally, because $C^{(1)}_{ijk0}$ vanishes for scalar modes, the last term in Eq.~(8) is
\begin{eqnarray}
&& f_{10}(\bar{\varphi})\epsilon^{\mu\nu\rho\sigma}
\bar{g}^{\kappa\eta}\bar{g}^{\lambda\zeta}C^{(1)}_{\mu\nu\eta\zeta}C^{(1)}_{\rho\sigma\kappa\lambda}
\nonumber\\&&=f_{10}(\bar{\varphi})\epsilon^{ijk0}\Big[4a^{-4}C^{(1)}_{ijlm}C^{(1)}_{k0lm}-8C^{(1)}_{ijl0}C^{(1)}_{k0l0}\Big]=0\;.
\end{eqnarray}
{\em We conclude that the leading corrections to the Gaussian correlations  of ${\cal R}$ are solely of the $k$-inflation type}.  This justifies the calculation of the effective Lagrangian for Gaussian scalar correlations in slow roll inflation in Section 3 of reference 3 even for generic theories of inflation.  In such theories the terms in Eq.~(3) that are left out in reference 3 can indeed be omitted in calculating the part of the effective Lagrangian quadratic in scalar fluctuations, {\em not because they are small}, but because as we have seen for scalar Gaussian fluctuations they yield nothing new.  But this is not the case for Gaussian tensor fluctuations, and does not seem to be the case when non-Gaussian correlations are considered.

We have so far only considered the terms in the effective action of second order in ${\cal R}$, which are needed to calculate Gaussian correlations, but for actions of the $k$-inflation type, which only involve first derivatives of fields, it is not difficult also to calculate terms in the action of higher order in ${\cal R}$, which generate non-Gaussian correlations.  For this purpose it is convenient to adopt a gauge in which there are no scalar perturbations to $g_{ij}$; that is, in which $g_{ij}=a^2(t)[\exp(D({\bf x},t))]_{ij}$, where $D_{ij}$ is a gravitational wave amplitude with $D_{ii}=0$ and $\partial_iD_{ij}=0$.  In this gauge, Eq.~(10) gives ${\cal R}=-H\delta\varphi/\dot{\bar{\varphi}}$.  If we tentatively ignore the interaction of the inflaton with gravitational perturbations, and assume that $H$, $f_1$, and $\dot{\bar{\varphi}}$ are varying slowly, then it is trivial, by simply setting $\varphi$ equal to $\bar{\varphi}+\delta\varphi$ in ${\cal L}$, and using $\dot{H}=-\dot{\bar{\varphi}}^2(M^2+4f_1\dot{\bar{\varphi}}^2)/2M_P^2$, to write a Lagrangian for $\pi\equiv -{\cal R}/H=\delta\varphi/\dot{\bar{\varphi}}$:  
\begin{eqnarray}
&&\sqrt{g}\Bigg[-\frac{M^2}{2}g^{\mu\nu}\partial_\mu\varphi \partial_\nu\varphi-M_P^2 U(\varphi)+
f_1(\varphi)(g^{\mu\nu}\partial_\mu\varphi \partial_\nu\varphi)^2\Bigg]\nonumber\\&&=\bar{\cal L}+a^3M_P^2\dot{H}
\left(-\dot{\pi}^2+a^{-2}(\vec{\nabla}\pi)^2\right)\nonumber\\&&+\frac{16a^3 M_P^4\dot{H}^2f_1(\bar{\varphi})}{M^4}\left(\dot{\pi}^2+\dot{\pi}^3-\frac{\dot{\pi}(\vec{\nabla}\pi)^2}{a^2}+\frac{\dot{\pi}^4}{4}-\frac{\dot{\pi}^2(\vec{\nabla}\pi)^2}{2a^2}+\frac{(\vec{\nabla}\pi)^4}{4a^2}\right)\;.~~~~~~
\end{eqnarray}
This agrees with the result obtained in Eq.~(28) of [3], except that here we include terms quartic in $\pi$.  In [3] the neglect of interactions of the inflaton with gravitational perturbations is justified on the basis of a ``high energy'' approximation, which amounts to the usual slow-roll approximation that $\epsilon\ll 1$, plus the assumption that, in our terms, $M^2\gg \epsilon H M_P$, which is much weaker than the assumption $M\gg \sqrt{2\epsilon }M_P$ that we found necessary to treat generic theories of inflation by the methods of effective field theory.  We see that the non-quadratic terms in ${\cal L}_0$ that can generate non-Gaussian correlations are suppressed in the slow roll approximation, as found by Maldacena[13], but in $\Delta L$ the coefficients of the quadratic and higher order terms are of the same order of magnitude.

\begin{center}
{\bf III. Tensor Fluctuations}
\end{center}

Tensor fluctuations appear solely in  the perturbation to the purely spatial metric:
\begin{equation}
h_{ij}({\bf x},t)=a^2(t)\Big[\exp D\Big]_{ij}({\bf x},t)\;,~~~~~D_{ii}=0\;,~~~~\partial_i D_{ij}=0\;,
\end{equation}
with $\delta\varphi=0$.  The first term in Eq.~(8) involves only the metric components $g^{00}$ and ${\rm Det}g_{\mu\nu}$, so it gets no contribution from tensor fluctuations.  On the other hand, here the second and third terms in (8) make a non-trivial contribution to the Lagrangian for $D_{ij}$.  Another straightforward calculation (dropping total derivatives) gives these terms as
\begin{eqnarray}
&&a^3f_9(\bar{\varphi})\bar{g}^{\mu\kappa}\bar{g}^{\nu\lambda}\bar{g}^{\rho\eta}\bar{g}^{\sigma\zeta}
C^{(1)}_{\kappa\lambda\eta\zeta}C^{(1)}_{\mu\nu\rho\sigma}\nonumber\\&&
=f_9(\bar{\varphi})\left[a^{-5}C^{(1)}_{ijkl}C^{(1)}_{ijkl}-4a^{-3}C^{(1)}_{ijk0}C^{(1)}_{ijk0}+4a^{-1}C^{(1)}_{i0k0}C^{(1)}_{i0k0}\right]\nonumber\\&&=
a^3f_9(\bar{\varphi})\Bigg\{\dot{D}_{ik}\Big[2H^2+2\nabla^2/a^2]\dot{D}_{ik}-4H\dot{D}_{ik}(\nabla^2/a^2)D_{ik}\nonumber\\&&~~~~~~~+2D_{ik}(\nabla^4/a^4)D_{ik}\Bigg\}\;,
\end{eqnarray}
and
\begin{eqnarray}
&&f_{10}(\bar{\varphi})\epsilon^{\mu\nu\rho\sigma}
\bar{g}^{\kappa\eta}\bar{g}^{\lambda\zeta}C^{(1)}_{\mu\nu\eta\zeta}C^{(1)}_{\rho\sigma\kappa\lambda}
=f_{10}(\bar{\varphi})\epsilon^{ijk0}\left[4a^{-4}C^{(1)}_{lmij}
C^{(1)}_{lmk0}-8a^{-2}C^{(1)}_{l0ij}C^{(1)}_{l0k0}\right]\nonumber\\&&=~~~~
4f_{10}(\bar{\varphi})\epsilon^{ijk0}\frac{\partial}{\partial t}\Big[D_{il}\partial_j\nabla^2D_{kl}\Big]\;.
\end{eqnarray}
The field equation for the tensor mode (with the term proportional to $f_9$ dropped for simplicity) is then
\begin{equation}
\ddot{D}_{il}+3H\dot{D}_{il}-(\nabla^2/a^2)D_{il}=-64\pi G \dot{f}_{10}a^{-3}\Big(\epsilon^{ijk0}\partial_j\nabla^2D_{kl}
+\epsilon^{ljk0}\partial_j\nabla^2D_{ki}\Big)\;.
\end{equation}
For a plane wave with co-moving wave number $\vec{k}$ in the 3-direction,
the only non-vanishing tensor amplitudes are $D_{11}=-D_{22}$ and $D_{12}=D_{21}$.  They satisfy the field equations
\begin{equation}
\ddot{D}_\pm+3H\dot{D}_\pm+(k^2/a^2)D_\pm=\mp 128\pi G(k/a)^3 \dot{f}_{10} D_\pm
\end{equation}
where $D_\pm\equiv D_{11}\mp i D_{12}$ are the amplitudes with helicity $\pm 2$.  As found in ref. [11], the wave equation depends on helicity because parity is violated.

\begin{center}
{\bf IV. A Non-Generic Example: Ghost Inflation}
\end{center}

Up to now, we have been concerned with generic theories of inflation, in which the dependence of the action on the inflaton field is unconstrained, and in consequence the characteristic mass $M$ cannot be taken to be much less than $\sqrt{\epsilon}M_P$.  For an  example of a different sort, we might impose on the action a shift symmetry, under a transformation $\varphi\rightarrow\varphi+$constant, which requires that the Lagrangian density involve only spacetime derivatives of $\varphi$ rather than $ \varphi$ itself.  This possibility was discussed briefly in [9], and in more detail under the name ``ghost inflation'' in [14].  We will take $\varphi$ to be normalized so that $\partial_\mu \varphi$ is dimensionless, and has an unperurturbed value at horizon exit of order unity.    The term in the Lagrangian density that depends only on $\partial_\mu\varphi$ is then
\begin{equation}
{\cal L}_0=  M^4\sqrt{g}P( -\partial_\mu\varphi\partial^\mu\varphi)\;,
\end{equation}
where $P(X)$ is a power series in $X$, with coefficients assumed to be of order unity, and $M$ is the characteristic mass of the theory.  Since powers of $\varphi$ are excluded by the shift symmetry, $M$ here can be much smaller than in generic theories of inflation, and in particular we will assume that $M$ is much less than the Planck mass $M_P$.  Any additional derivatives acting on $\partial_\mu\varphi$ or on the metric yield factors of order $H\sim M^2/M_P\ll M$, so Eq.~(25) along with the Einstein term can be taken as the leading term in ${\cal L}$, with any correction terms suppressed by factors of $H/M$.  

Let us first consider a theory in which (25) is the whole Lagrangian density for the scalar field, with no higher-derivative corrections.  The field equation for the unperturbed scalar field $\bar{\varphi}(t)$ in this theory is
\begin{equation}
\frac{d}{dt}\Big(a^3P'(\dot{\bar{\varphi}}^2)\dot{\bar{\varphi}}\Big)=0\;.
\end{equation}
As noted in [9], in the limit of late time when $a\rightarrow\infty$, either $\dot{\bar{\varphi}}\rightarrow 0$, or 
$\dot{\bar{\varphi}}\rightarrow v$, where $v$ is a quantity of order unity satisfying $P'(v^2)=0$.  We will consider only the latter case.  In ref. [14] the limit $\bar{\varphi}=v t$ is supposed to be already reached, in which case interesting fluctuations occur only when higher-derivative correction terms are added to (25).  But if we take $\bar{\varphi}(t)$ to be only close to $v t$, but not yet there, then we find a non-trivial spectrum of propagating fluctuations even when no correction terms are added to (25). In this case the solution of Eq.~(26) (with an appropriate normalization of $a(t)$) has $\dot{\bar{\varphi}}\rightarrow v+ a^{-3}$; the speed of sound is $c_s\rightarrow \sqrt{1/va^3}$; the expansion rate approaches a limit $H_\infty=(M^2/M_P)\sqrt{P(v^2)/3}$; and the Fourier transform of ${\cal R}$ is
\begin{equation}
{\cal R}_k\propto a^{-3/2}H_{3/5}^{(1)}\left(\frac{2k}{5H_\infty v^{1/2} a^{5/2}}\right)\;,
\end{equation}
with a $k$-independent constant of proportionality.  At late times, when the perturbation wave length is outside the acoustic horizon, this approaches a time-independent quantity ${\cal R}_k^o$, with
\begin{equation}
|{\cal R}^o_k|^2\propto k^{-6/5}\;,
\end{equation}
corresponding to a conventional scalar slope index $n_S=14/5$, which of course is empirically ruled out.  Thus to have a realistic theory of this sort, we must consider corrections to the leading term (25).


The first correction to this Lagrangian density contains just one factor of a second derivative of $\varphi$, and in general is of the form
\begin{equation}
\Delta {\cal L}=M^3\sqrt{g}Q(-\partial_\mu\varphi\partial^\mu\varphi)\Box\varphi
\;,
\end{equation}
and is therefore suppressed relative to (25) by factors of order $H/M\approx M/M_P$.  
(A term proportional to $g^{\rho\kappa}g^{\sigma\lambda}\varphi_{,\kappa}\varphi_{,\lambda}
\varphi_{,\rho;\sigma}$ can be put in the form (29) by adding suitable total derivatives.  In ref. [14] these terms were excluded by imposing an additional symmetry under the reflection $\varphi\rightarrow-\varphi$, in which case the first correction is quadratic rather than linear in second derivatives of $\varphi$.)  Once again, we must eliminate the second time derivatives in $\Delta{\cal L}$ by setting $\ddot{\varphi}$  equal to the same quantity as given by the field equation derived from the leading part of the Lagrangian
\begin{equation}
P'(-\partial_\nu\varphi\partial^\nu\varphi)\Box\varphi-2P''(-\partial_\nu\varphi\partial^\nu\varphi)(\partial_\nu\varphi)_{;\mu}\partial^\mu\varphi
\partial^\nu\varphi=0\;.
\end{equation}
This is pretty complicated, so for simplicity let us consider the case of a metric fixed in the flat-space Robertson--Walker form.  Then after using (30) to eliminate second time derivatives in (29), the correction term is
\begin{eqnarray}
\Delta {\cal L}&=&M^3a^3\left (\frac{2Q P''}{P'+2P''\dot{\varphi}^2}\right)\,\Big(-2a^{-2}\dot{\varphi}\partial_i\varphi\partial_i\dot{\varphi}\nonumber\\&&
+a^{-2}H\dot{\varphi}\partial_i\varphi\partial_i\varphi
+a^{-4}\partial_i\varphi\partial_j\varphi\partial_i\partial_j\varphi+3H\dot{\varphi}^3-a^{-2}\dot{\varphi}^2\nabla^2\varphi\Big)\;,
\end{eqnarray}
where $Q$, $P'$ and $P''$ all have arguments
$-\partial_\mu\varphi\partial^\mu\varphi=\dot{\varphi}^2-a^{-2}\partial_i\varphi\partial_i\varphi$.  This is the correction that has to be added to ${\cal L}_0$ in order to find the commutation relations of the field as well as the field equations by canonical quantization.

\vspace{12pt}

I am grateful for discussions with N. Arkani-Hamed, J. Distler, J. Meyers, S. Odintsov, S. Paban, D. Robbins, and L. Senatore.  This material is based upon work supported by the National Science Foundation under Grant No. PHY-0455649.

\begin{center}
{\bf References}
\end{center}

\begin{enumerate}
\item For  reviews with references to the original literature, see V. Mukhanov, {\em Physical Foundations of Cosmology} (Cambridge University Press, 2005); S. Weinberg, {\em Cosmology} (Oxford University Press, 2008).

\item This is based on  third year WMAP results; D. Spergel {\em et al.}, Astrophys. J. Suppl. {\bf 170}, 288 (2007).

\item This is different from the approach followed in an interesting recent paper on the effective field theory of inflation by C. Cheung, P. Creminelli, A. L. Fitzpatrick, J. Kaplan, and L. Senatore, arxiv:0709.0293, whose calculations do not include any of the terms in Eq.~(3) following the first term.  Their paper does not spell out the rules governing which terms are to be included in the corrections of leading order, but on the basis of a private communication with Senatore, I gather that in judging how much various correction terms are suppressed, Cheung {\em et al.}  do not count spacetime derivatives acting on the background scalar or metric fields in a co-moving (or ``unitary'') gauge in which $\varphi$ equals its unperturbed value $\bar{\varphi}$, but only count derivatives acting on fluctuations in this gauge.  In co-moving gauge the first term in (3) is $\sqrt{g}f_1(\bar{\varphi})(g^{00}\dot{\bar{\varphi}}{}^2)^2$, and the factors of $\dot{\bar{\varphi}}$ are not counted as suppressing this term, because $|\dot{\bar{\varphi}}_c|$ is much larger than $H^2$.  But as we have seen, in generic theories of inflation the characteristic mass scale $M$ of the effective field theory must be at least as large as $\sqrt{\epsilon}M_P$, and  the  quantity $|\dot{\bar{\varphi}}_c|/M$ is then no larger than $H$, so that the first term in (3) is no less suppressed than the other terms.   
(Cheung {\em et al.} do at first include terms involving the extrinsic curvature of the spacelike surface with $\varphi$ constant, but later drop these terms.  The extrinsic curvature is not included here in Eq.~(3), because in a general gauge it does not give a local term in the action.  But Cheung {\em et al.} stick to co-moving gauge, in which the extrinsic curvature can be expanded in a series of local functions,  and these do yield some though not all of the terms in Eq.~(3).)  The approach of Cheung {\em et al.} is justified in theories with a much smaller value of $M$ than considered here, which is possible if the dependence of the action on the inflaton field is limited by symmetry  principles or other consequences of an underlying theory.  This case is discussed briefly at the end of the present paper.                                                                                                                                                                                                                                                                                                                                                                                                                                                                                                                                                                                                                                                                                                                                                                                                                                                                                                                                                                                                                                                                                                                                                                                                                                                                                                                                                                                                                                                                                                                                                                        

\item The same list of terms with just four spacetime derivatives (aside from the term that violates parity conservation) has been given   by E. Elizalde, A. Jacksenaev, S. D. Odintsov, and I. L. Shapiro, Phys. Lett. B {\bf 328}, 297 (1994); Class. Quant. Grav. {\bf 12}, 1385 (1995), but not in the context of effective field theory.  (In the original preprint of the present paper,  written before I had seen the work of Elizalde {\em et al.}, this list contained an additional term proportional to $\sqrt{g}R^{\mu\nu}\varphi_{,\mu;\nu}$.  This term is redundant, because by using the Bianchi identity and discarding  total derivatives it may be expressed as a linear combination of the fourth, fifth, and sixth terms listed here in Eq.~(3).)

\item M. Ostrogradski, Mem. Act. St. Petersbourg {\bf VI 4}, 385 (1850).  For a modern account, see F. J. de Urries and J. Julve, J. Phys. {\bf A31}, 6949 (1998).

\item R. Kallosh, J. U. Kang, A. Linde, and V. Mukhanov, arXiv:0712.2040.

\item R. S. Arnowitt, S. Deser, and C. W. Misner, in {\em Gravitation: An Introduction to Current Research}, edited by L. Witten (Wiley, New York, 1962).

\item See, {\em e. g.}, J. Z. Simon, Phys. Rev. D {\bf 41}, 3729 (1990).

\item C. Armend\'{a}riz-Pic\'{o}n, T. Damour, and V. F. Mukhanov, Phys. Lett. B {\bf 458}, 209 (1999).

\item  S. Kawai, M. Sakagami, and J. Soda, Phys. Lett. B {\bf 437}, 284 (1998); S. Kawai and J. Soda, Phys. Lett. B {\bf 460}, 41 (1999); S. Nojiri, S. D. Odintsov, and M. Sasaki, Phys. Rev. D {\bf 71}, 123509 (2005); G. Calcagni, S. Tsujikawa, and M. Sami, Class. Quant. Grav. {\bf 22}, 3977 (2005); G. Calcagne, B. de Carlos, A. De Felice, Nucl. Phys. B {\bf 752}, 404 (2006); I. P. Neupane and B. M. N. Carter, J. Cosm. \& Astropart. Phys. 06, 004 (2006):  G. Cognola, E. Elizalde, S. Nojiri, S. D. Odintsov, and S. Zerbini, Phys. Rev. {\bf D73}, 084007 (2006); B. Leith and I. P. Neupane, J. Cosm. \& Astropart. Phys. 05, 019 (2007); S. Tsujikawa and M. Sami, J. Cosm. \& Astropart. Phys. 07, 006 (2007); K. Bamba, Z-K Guo, and N. Ohta, arxiv:0707.4334.  Some of these articles deal with instabilities produced by a Gauss--Bonnet term, but no instability arises if this term is treated as a correction term in an effective field theory.  The first term in Eq.~(6) along with a Gauss--Bonnet term equivalent to the second term in Eq.~(6) were encountered in a low-energy limit of string theory by Z. K. Guo, N. Ohta, and S. Tsujikawa, Phys. Rev. D {\bf 75}, 023520 (2007).

\item There is a large literature on a parity violating term  of this form.  See, for instance,  A. Lue, L. Wang, and M. Kamionkowski, Phys. Rev. Lett. {\bf 83}, 156 (1999);  S-Y. Pi and R. Jackiw, Phys. Rev. {\bf D68}, 104012 (2003); M. Satoh, S. Kanno, and J. Soda, Phys. Rev. {\bf D77}, 023526 (2008).  Leptogenesis due to this term was considered by S. H. S. Alexander, M. E. Peskin, and M. M. Sheikh-Jabbari, Phys. Rev. Lett. {\bf 96}, 081301 (2006).

\item J. Garriga and V. F. Mukhanov, Phys. Lett. B {\bf 458}, 219 (1999).

\item J. Maldacena, JHEP {\bf 0305}, 013 (2003).

\item  N. Arkani-Hamed, P. Creminelli, S. Mukohyama, and M. Zaldarriaga, JCAP {\bf 04}, 001 (2001); N. Arkani-Hamed, H-C Cheng, M. A. Luty, and S. Mukohyama, JHEP {\bf 05}, 074 (2004); S. Mukohyama, JCAP {\bf 0610}, 011 (2006).

\end{enumerate}

\end{document}